\documentclass[twoside,fleqn]{article}
\usepackage{graphicx}
\usepackage{espcrc2}
\usepackage{psfig}

\def\ed{\end{document}}

\def\rp{$R_p \hspace{-1em}/\;\:$ }
\def\Table#1{Table~\ref{#1}}
\def\bold#1{\setbox0=\hbox{$#1$} 
     \kern-.025em\copy0\kern-\wd0 
     \kern.05em\copy0\kern-\wd0 
     \kern-.025em\raise.0433em\box0 }
\def\vev#1{\left\langle #1\right\rangle}
\def\beq{\begin{equation}}
\def\eeq{\end{equation}}
\def\ba{\begin{array}}
\def\ea{\end{array}}
\def\Slash#1{#1\!\!\!\! /}
\def\beqa{\begin{eqnarray}}
\def\eeqa{\end{eqnarray}}
\def\vb#1{\vbox to #1 pt{}}

\def\ds{\displaystyle}
\def\21{$SU(2) \otimes U(1)$}
\def\cimaum#1{\hbox{\raise .5ex\hbox{$#1$}}}

\def\ifmath#1{\relax\ifmmode #1\else $#1$\fi}
\def\half{\ifmath{{\textstyle{1 \over 2}}}}
\def\quarter{\ifmath{{\textstyle{1 \over 4}}}}
\def\eighth{\ifmath{{\textstyle{1 \over 8}}}}

\def\npb#1#2#3{{\it Nucl.\ Phys.\ }{\bf B #1} (#2) #3}
\def\plb#1#2#3{{\it Phys.\ Lett.\ }{\bf B #1} (#2) #3}
\def\prd#1#2#3{{\it Phys.\ Rev.\ }{\bf D #1} (#2) #3}
\def\prep#1#2#3{{\it Phys.\ Rep.\ }{\bf #1} (#2) #3}
\def\prl#1#2#3{{\it Phys.\ Rev.\ Lett.\ }{\bf #1} (#2) #3}

\def\jetpl#1#2#3{{\it Sov.\ Phys.\ JETP Lett.\ }{\bf #1} (#2) #3}
\def\rnc#1#2#3{{\it Riv. Nuovo Cimento }{\bf #1} (#2) #3}

\def\hepph#1{{\tt hep-ph/#1}}

\newcommand {\ignore}[1]{}

\newcommand{\AmS}{{\protect\the\textfont2
  A\kern-.1667em\lower.5ex\hbox{M}\kern-.125emS}}

\title{Phenomenology of Supersymmetric Theories with and without 
R-Parity}

\author{J. C. Rom\~ao\address{Instituto Superior T\'ecnico, 
Departamento de F\'{\i}sica\\
A. Rovisco Pais 1, 1049-1001 Lisboa, Portugal}
        \thanks{Talk given at the
EuroConference on Frontiers in Particle Astrophysics and Cosmology, 
San Feliu de Guixols, Spain, 30 September - 5 October, 2000.}
        \thanks{This work was supported by the TMR network grant
ERBFMRXCT960090 of the European Union.}
}

\begin{document}

\begin{abstract}
We review supersymmetry models with and without R-parity. After
briefly describing the Minimal Supersymetric Standard Model and its
particle content we move to models where R--parity is broken, 
either spontaneously or explicitly. In this last case we consider 
the situation where R--parity is broken via bilinear terms in the 
superpotential. The radiative breaking of these models is described in
the context of $b$--$\tau$ and $b$--$\tau$--$t$ unification.
Finally we review the phenomenology of these R-parity violating models.
\end{abstract}

\maketitle

\section{The Minimal Supersymmetric Standard Model}

\subsection{Motivation for Supersymmetry}

Although there is not yet direct experimental evidence for 
supersymmetry (SUSY)~\cite{SUSY}, there are many theoretical arguments 
indicating that SUSY might be of
relevance for physics below the 1 TeV scale. 
In fact SUSY interrelates matter fields (leptons and quarks) with 
force fields (gauge and/or Higgs bosons) and as
local SUSY implies gravity (supergravity) it could provide a way to
unify gravity with the other interactions.
As SUSY and supergravity have fewer divergences than conventional
field theories, the hope is that it could provide a consistent
(finite) quantum gravity theory. Finally and most important,
SUSY can help to understand the mass problem, in particular solve the
naturalness problem ( and in some models even the hierarchy problem)
if SUSY particles have masses $\le {\cal O} (1 \hbox{TeV})$.

\subsection{R--Parity}

In the discussions of SUSY phenomenology there is a quantum number
called {\it R-Parity} that plays an important role:
\beq
R_P=(-1)^{2J +3B +L}
\eeq
In the MSSM this quantity is conserved. This implies that 
SUSY particles are pair produced, every SUSY particle decays into
another SUSY particle and there is a {\it LSP} that it is stable 
($\Slash{E}$ signature).

\subsection{The Model}

The MSSM Lagrangian is specified~\cite{MSSM} by the R-parity conserving
superpotential $W$, 
\beqa
\label{W_MSSM}
W_{MSSM}&\hskip -2mm=\hskip -2mm&\varepsilon_{ab}\left[ 
 h_U^{ij}\widehat Q_i^a\widehat U_j\widehat H_2^b 
+h_D^{ij}\widehat Q_i^b\widehat D_j\widehat H_1^a \right. \cr 
&&\left. \hskip 10mm
+h_E^{ij}\widehat L_i^b\widehat R_j\widehat H_1^a 
-\mu\widehat H_1^a\widehat H_2^b \right] 
\eeqa
where $i,j=1,2,3$ are generation indices, $a,b=1,2$ are $SU(2)$ 
indices, and $\varepsilon$ is a completely antisymmetric $2\times2$  
matrix, with $\varepsilon_{12}=1$. To this we have to add the
SUSY soft breaking terms,
\begin{eqnarray} 
\label{SoftMSSM}
V_{soft}^{MSSM}\hskip -3mm&=&\hskip -3mm 
M_Q^{ij2}\widetilde Q^{a*}_i\widetilde Q^a_j \!+\! M_U^{ij2} 
\widetilde U^*_i\widetilde U_j\! +\! M_D^{ij2}\widetilde D^*_i 
\widetilde D_j \cr
\vb{15}
&&\hskip -5mm
+M_L^{ij2}\widetilde L^{a*}_i\widetilde L^a_j 
+M_R^{ij2}\widetilde R^*_i\widetilde R_j\cr
\vb{15}
&&\hskip -5mm
+ m_{H_1}^2 H^{a*}_1 H^a_1 + m_{H_2}^2 H^{a*}_2 H^a_2 \cr
\vb{15}
&&\hskip -5mm
-\! \left[\half M_s\lambda_s\lambda_s \!+\! \half M\lambda\lambda 
\!+\! \half M'\lambda'\lambda'\!+\! h.c.\right]\cr
\vb{15}
&&\hskip -5mm
+\varepsilon_{ab}\left[ 
A_U^{ij}h_U^{ij}\widetilde Q_i^a\widetilde U_j H_2^b 
+A_D^{ij}h_D^{ij}\widetilde Q_i^b\widetilde D_j H_1^a \right. \cr
\vb{15}
&&\left. \hskip 3mm
+A_E^{ij}h_E^{ij}\widetilde L_i^b\widetilde R_j H_1^a 
-B\mu H_1^a H_2^b \right]  
\end{eqnarray}

\noindent 
The electroweak symmetry is broken when the two Higgs doublets  
$H_1$ and $H_2$ acquire vevs

\beq
H_1=
\left(
\matrix{H^0_1\cr H^-_1}
\right)\quad ; \quad
H_2=
\left(
\matrix{H^+_2\cr H^0_2}
\right)
\eeq
where
$H^0_1={1\over{\sqrt{2}}}\left(\sigma^0_1+v_1+i\varphi^0_1\right)$, and
$H^0_2={1\over{\sqrt{2}}}\left(\sigma^0_2+v_2+ i\varphi^0_2\right)$. Our
definitions are such that
\beq
m_W^2=\quarter g^2v^2\quad ; \quad
v^2\equiv v_1^2+v_2^2=(246 \; \rm{GeV})^2
\eeq
The full scalar potential at tree level is then
\beq
V_{total}  = \sum_i \left| { \partial W \over \partial z_i} \right|^2 
	+ V_D + V_{soft} 
\eeq
The scalar potential contains linear terms 
\beq
V_{linear}=t_1^0\sigma^0_1+t_2^0\sigma^0_2
\eeq
where
\begin{eqnarray} 
\hskip-2mm
t_1\hskip -1mm&=&\hskip -1mm
m_1^2 v_1\!-\!B\mu v_2\!+\!
\eighth(g^2\!+\!g'^2)v_1(v_1^2\!-\!v_2^2)\,, 
\cr
\vb{15}
t_2\hskip -1mm&=&\hskip -1mm
m_2^2 v_2\!-\!B\mu v_1\!-\eighth(g^2
\!+\!g'^2)v_2(v_1^2\!-\!v_2^2) 
\end{eqnarray} 
and
$m_i^2=m_{H_i}^2\!+\!\mu^2$.
At the minimum one should have
\beq
t_i=0 \quad ; \quad i=1,2
\eeq
and $m_{H_2}^2 < 0$. Now two approaches are possible. In the first
the values of the parameters at the weak scale are completely arbitrary.
In the second the theory at weak scale is obtained from a GUT and
there are few parameters at GUT scale. This possibility is more
constrained (CMSSM). In the second approach one usually takes 
the N=1 SUGRA conditions:
\beqa
\label{SUGRA}
&&A_t = A_b = A_{\tau} \equiv A \:, \cr
&&\vb{15}
B=B_2=A-1 \:, \cr
&&\vb{15}
m_{H_1}^2 = m_{H_2}^2 = M_{L}^2 = M_{R}^2 = m_0^2 \:, \cr
&&\vb{15}
M_{Q}^2 =M_{U}^2 = M_{D}^2 = m_0^2 \:, \cr
&&\vb{15}
M_3 = M_2 = M_1 = M_{1/2} 
\eeqa
After using the minimization equations one ends up with three
independent parameters. These are normally taken to be 
$\tan\beta$ and the masses of two of the physical Higgs bosons.
It is remarkable that with so few parameters we can get
the correct values for the parameters, in particular $m_{H_2}^2 <0$.

\subsection{Particle Content}

The minimal particle content of the MSSM is described in 
Table \ref{particlecontent}.
\begin{table}
\caption{MSSM Particle Content}
\begin{center}
\begin{tabular}{|l|c|}\hline
Supermultiplet&$SU_c(3)\otimes SU_L(2)\otimes U_Y(1)$\cr
&Quantum Numbers\cr\hline\hline
\vb{14}
$\widehat V_1\equiv(\lambda',W_1^{\mu})$&$U_Y(1)$ Gauge multiplet \cr
\vb{14}
$\widehat V_2\equiv(\lambda^a,W_2^{\mu a})$&$SU_L(2)$ Gauge multiplet \cr
\vb{14}
$\widehat V_3\equiv(\widetilde{g}^b,W_3^{\mu b})$&$SU_c(3)$ 
Gauge multiplet \cr\hline 
\vb{14}
$\widehat L_i\equiv(\widetilde{L},L)_i$&$(1,2,-\half)$\cr
\vb{14}
$\widehat R_i\equiv(\widetilde{\ell}_R,\ell^c_L)_i$&$(1,1,1)$\cr\hline
\vb{14}
$\widehat Q_i\equiv(\widetilde{Q},Q)_i$&$(3,2,\frac{1}{6})$\cr
\vb{14}
$\widehat D_i\equiv(\widetilde{d}_R,d^c_L)_i$&$(3,1,\frac{1}{3})$\cr
\vb{14}
$\widehat U_i\equiv(\widetilde{u}_R,u^c_L)_i$&$(3,1,-\frac{2}{3})$\cr\hline
\vb{14}
$\widehat H_1\equiv(H_1,\widetilde{H}_1)$&$(1,2,-\frac{1}{2})$\cr
\vb{14}
$\widehat H_2\equiv(H_2,\widetilde{H}_2)$&$(1,2,+\frac{1}{2})$\cr\hline
\end{tabular}
\end{center}
\label{particlecontent} 
\end{table}
From this table one can see that the MSSM more than {\it doubles} the 
SM particle content. 
For SUSY to be relevant for the hierarchy problem these SUSY partners
have to have masses below the 1 TeV scale. Then they should be seen at
LEP and/or LHC.

\subsection{The Higgs Mass}

\subsubsection{Tree Level}

The tree level mass matrices are

\beqa
\bold{M_R}^2_{TL}&=&\left(
\matrix{
\cot \beta & -1\cr
-1 & \tan \beta\cr}\right)\,
\half m_Z^2 \sin 2 \beta \cr
\vb{15}
&&
+ 
\left( \matrix{
\tan \beta & -1 \cr
-1 &\cot \beta\cr}\right)\,
\Delta_{TL}
\eeqa
and
\beq
\bold{M_I}^2_{TL}=\left(
\matrix{
\tan \beta & -1 \cr
\vb{16}
-1 &\cot \beta\cr}\right)\,
\Delta_{TL} 
\eeq
where
\beq
\Delta_{TL}=B\mu
\eeq
From these we obtain the masses,
\beqa
m_A^2
\hskip -3mm&=&\hskip -3mm
2\Delta_{TL} / \sin 2 \beta\cr
\vb{15}
m_{h,H}^2
\hskip -3mm&=&\hskip -3mm
\half\left[\vb{13} m_A^2 +m_Z^2  \right. \cr 
&&\left.\hskip -2mm \mp
\sqrt{(m_A^2+m_Z^2)^2 -4 m_A^2 m_Z^2 \cos^2 2 \beta}\right]
\eeqa
with 
\beq
m_h^2+m_H^2=m_A^2+m_Z^2
\eeq
and the very important result
\beq
\label{TreeLevelBound}
m_h<m_A<m_H\quad ;\quad m_h<m_Z<m_H
\eeq

\subsubsection{Radiative Corrections}

The tree level bound of Eq.~(\ref{TreeLevelBound}) is in fact evaded
because the radiative corrections due to the top mass are quite 
large. The mass matrices are now, 
\beqa
\bold{M_R}^2_{1L}
\hskip -2mm&=&\hskip -2mm
\bold{M_R}^2_{TL} + \
\left( \matrix{
\tan \beta & -1 \cr
-1 &\cot \beta\cr}\right)\,
\Delta_{1L} \cr
\vb{18}
&&
+ \frac{3g^2}{16\pi^2m_W^2}\, 
\left( \matrix{
\Delta_{11} & \Delta_{12} \cr
\Delta_{21} & \Delta_{22}\cr}\right)\,
\eeqa
and
\beq
\bold{M_I}^2=\left(
\matrix{
\tan \beta & -1 \cr
-1 &\cot \beta\cr}\right)\,
(\Delta_{TL}+\Delta_{1L}) 
\eeq
The $\Delta_{ij}$ are complicated expressions. The
most important one is
\beq
\Delta_{22}=\frac{m_t^4}{\sin^2 \beta}\,
\log \left(\frac{m^2_{\tilde{t}_1}m^2_{\tilde{t}_2}}{m_t^4}\right)
\label{Delta22}
\eeq
Due to the strong dependence on the top mass 
the CP--even states are the most affected. The mass of the lightest
Higgs boson, $h$ can now be as large as $140\ GeV$.

\section{Spontaneously Broken R-Parity}

\subsection{The Original Proposal}

In the original proposal~\cite{OriginalSBRP} the content was just the 
MSSM and the breaking was induced by
\beq
\vev{\tilde{\nu}_{\tau}} = v_L
\eeq
The problem with this model was that the Majoron $J$ coupled to $Z^0$ with 
gauge strength and therefore the decay
$Z^0 \rightarrow \rho_L J$ contributed to the invisible $Z$ width the
equivalent of half a (light) neutrino family. After LEP I this was
excluded.

\subsection{A Viable Model for SBRP}

The way to avoid the previous difficulty 
is to enlarge the model and make $J$ mostly out of {\it
isosinglets}. This was proposed by Masiero and Valle~\cite{MV}. The
content is the MSSM plus a few Isosinglet Superfields that carry
lepton number,
\beq
\nu^c_i\equiv (1,0,-1)\ ; \ S_i \equiv(1,0,1) \ ; \
\Phi\equiv (1,0,0)
\eeq
The model is defined by the superpotential \cite{MV,SBRpV},
\beqa
W&=&h_u u^c Q H_u + h_d d^c Q H_d + h_e e^c L H_d \cr
\vb{18}
&&+(h_0 H_u H_d - \mu^2 ) \Phi \cr
\vb{18}
&&+ h_{\nu} \nu^c L H_u + h \Phi \nu^c S
\eeqa
where the lepton number assignments are shown in \Table{table0}.
\begin{table}
\caption{Lepton number assignments.}
\begin{tabular}{lccccc} \hline
Field & $L$ & $e^c$ & $\nu^c$ & $S$ & others  \\ 
Lepton \# & $1$ & $-1$ & $-1$& $1$ & $0$ \\ \hline
\end{tabular}
\label{table0}
\end{table}
The spontaneous breaking of R parity
and lepton number is driven by \cite{SBRpV}
\beq
v_R = \vev {\tilde{\nu}_{R\tau}} \quad
v_S = \vev {\tilde{S}_{\tau}} \quad
v_L = \vev {\tilde{\nu}_{\tau}}
\eeq
The electroweak breaking and fermion masses arise from
\beq
\vev {H_u} = v_u ~~~~~
\vev {H_d} = v_d
\eeq
with $v^2 = v_u^2 + v_d^2$ fixed by the W mass.
The Majoron is given by the imaginary part of 
\beq
\frac{v_L^2}{V v^2} (v_u H_u - v_d H_d) +
              \frac{v_L}{V} \tilde{\nu_{\tau}} -
              \frac{v_R}{V} \tilde{{\nu^c}_{\tau}} +
              \frac{v_S}{V} \tilde{S_{\tau}}
\eeq
where $V = \sqrt{v_R^2 + v_S^2}$. 
Since the Majoron
is mainly an \21 singlet it does not contribute to the
invisible $Z^0$ decay width.

\subsection{Some Results on SBRP}

The SBRP model has been extensively studied. The implications for
accelerator \cite{SBRPAcc} and non--accelerator \cite{SBRPNonAcc}
physics have been presented  before and we will not discuss them here
\cite{brasil98}. As some of the recent work that we will describe at
the end of this talk has to do with the neutrino properties in the 
context of $R_P$ models we will only review here the neutrino results. 

\begin{itemize}
\item
{\it Neutrinos have mass.}
Neutrinos are massless at Lagrangian level but get mass from the
mixing with neutralinos\cite{paulo,npb}. In the SBRP model it is
possible to have non zero masses for two neutrinos \cite{npb}.
\item
{\it Neutrinos mix.}
The mixing is related to the the 
coupling matrix $h_{\nu_{ij}}$. This matrix  has to be non diagonal in
generation space to allow
\beq
\nu_{\tau} \rightarrow \nu_{\mu} + J 
\eeq
and therefore evading~\cite{npb} the {\it Critical Density Argument} against
$\nu's$ in the MeV range. 
\item
{\it Avoiding BBN constraints on the $m_{\nu_{\tau}}$.}
In the {\it SM} BBN arguments~\cite{bbnothers} rule out $\nu_{\tau}$ 
masses in the range
\beq
0.5\ MeV < m_{\nu_{\tau}} < 35 MeV
\eeq
We have shown~\cite{bbnpaper} that {\it SBRP} models can evade that 
constraint due to new annihilation channels
\beq
\nu_{\tau} \nu_{\tau} \rightarrow J J 
\eeq
\end{itemize}

\section{Explicitly Broken R-Parity}

The most general superpotential $W$ with the particle content
of the MSSM is given by \cite{e3others,chaHiggsEps}

\beq
\label{WRpTotal}
W= W_{MSSM} + W_{\Slash{R}}
\eeq
where $W_{MSSM}$ is given in Eq.~\ref{W_MSSM} and
\begin{eqnarray}
W_{\Slash{R}}&=&\varepsilon_{ab}\left[
 \lambda_{ijk}\widehat L_i^a\widehat L_j^b\widehat R_k
+\lambda^{'}_{ijk}\widehat D_i \widehat L_j^a\widehat Q_k^b \right]\cr
&&\cr
&& 
+\lambda^{''}_{ijk} \widehat D_i \widehat D_j\widehat U_k 
+ \varepsilon_{ab}\, 
\epsilon_i\widehat L_i^a\widehat H_u^b 
\label{WRPV}
\end{eqnarray}
The set of soft supersymmetry
breaking terms are
\beq
V^{soft}=V^{soft}_{ MSSM}+V^{soft}_{\Slash{R}}
\eeq
where $V^{soft}_{ MSSM}$ are given in Eq.~\ref{SoftMSSM} and
\begin{eqnarray}
V^{soft}_{\Slash{R}}&\hskip -3mm=\hskip -3mm&
\varepsilon_{ab}\left[
A_{\lambda}^{ij}\widetilde L^a_i\widetilde L_j^b \widetilde R_k
+A_{\lambda'}^{ijk}\widetilde D_i\widetilde L_j^a \widetilde Q_k^b 
\right] \cr
&\hskip -3mm\hskip -3mm&
\vb{18}
+A_{\lambda^{''}}^{ij}\widetilde D_i\widetilde D_j \widetilde U_k 
+\varepsilon_{ab}\,
B_i\epsilon_i\widetilde L^a_i H_u^b + h.c.
\label{SoftRPV}
\end{eqnarray}

\section{Bilinear R-Parity Violation}

\subsection{The Model}

The superpotential $W$ for the bilinear $R_P$ violation model is 
obtained from Eq.~(\ref{WRpTotal}) by putting to
zero~\cite{e3others,chaHiggsEps}  all the
trilinear couplings,
\beq
W=W_{MSSM} + \varepsilon_{ab}\ \epsilon_i\ \widehat L_i^a\widehat H_2^b
\eeq
The set of soft supersymmetry
breaking terms are
\beq
V_{soft}=V_{soft}^{MSSM} + \varepsilon_{ab}\
B_i\ \epsilon_i\ \widetilde L^a_i H_2^b
\eeq
The electroweak symmetry is broken when the VEVS of 
the two Higgs doublets $H_d$
and $H_u$, and the sneutrinos.
\begin{eqnarray}
H_d&=&{{{1\over{\sqrt{2}}}[\chi^0_d+v_d+i\varphi^0_d]}\choose{
H^-_d}} \\
H_u&=&{{H^+_u}\choose{{1\over{\sqrt{2}}}[\chi^0_u+v_u+
i\varphi^0_u]}}\\
L_i&=&{{{1\over{\sqrt{2}}}
[\tilde\nu^R_{i}+v_i+i\tilde\nu^I_{i}]}\choose{\tilde\ell^{i}}}
\end{eqnarray}
The gauge bosons $W$ and $Z$ acquire masses
\beq
m_W^2=\quarter g^2v^2 \quad ; \quad m_Z^2=\quarter(g^2+g'^2)v^2
\eeq
where
\beq
v^2\equiv v_d^2+v_u^2+v_1^2+v_2^2+v_3^2=(246 \; {\rm GeV})^2
\eeq

\noindent
The bilinear
R-parity violating term {\sl cannot} be eliminated by superfield
redefinition~\cite{marco-2}.
The reason is that the bottom Yukawa coupling, usually neglected,
plays a crucial role in splitting
the soft-breaking parameters $B$ and $B_2$ as well as the scalar
masses $m_{H_1}^2$ and $M_L^{2}$, assumed to be equal at the
unification scale.

\noindent
The full scalar potential may be written as

\beq
V_{total}  = \sum_i \left| { \partial W \over \partial z_i} \right|^2
+ V_D + V_{soft} + V_{RC}
\eeq
where $z_i$ denotes any one of the scalar fields in the
theory, $V_D$ are the usual $D$-terms, $V_{soft}$ the SUSY soft
breaking terms, and $V_{RC}$ are the 
one-loop radiative corrections. 

\noindent
In writing $V_{RC}$ we  use the diagrammatic method and find 
the minimization conditions by correcting to one--loop the tadpole
equations. 
This method has advantages with respect to the effective potential when
we calculate the one--loop corrected scalar masses.
The scalar potential contains linear terms
\beq
V_{linear}=t_d\sigma^0_d+t_u\sigma^0_u+t_i\tilde\nu^R_{i}
\equiv t_{\alpha}\sigma^0_{\alpha}\,,
\eeq
where we have introduced the notation
\beq
\sigma^0_{\alpha}=(\sigma^0_d,\sigma^0_u,\nu^R_1,\nu^R_2,\nu^R_3)
\eeq
and $\alpha=d,u,1,2,3$. The one loop tadpoles are
\begin{eqnarray}
t_{\alpha}&=&t^0_{\alpha} -\delta t^{\overline{MS}}_{\alpha}
+T_{\alpha}(Q)\cr
\vb{22}
&=&t^0_{\alpha} +T^{\overline{MS}} _{\alpha}(Q)
\label{tadpoles}
\end{eqnarray}
where $T^{\overline{MS}} _{\alpha}(Q)\equiv -\delta t^{\overline{MS}}_{\alpha}
+T_{\alpha}(Q)$ are the finite one--loop tadpoles.

\subsection{Main Features}

The $\epsilon$--model is a one (three) parameter(s) generalization of
the MSSM.
It can be thought as an effective model showing the more important 
features of the SBRP--model~\cite{SBRpV} at the weak scale. 
The mass matrices, charged and neutral currents, are similar to the
SBRP--model if we identify
\beq
\epsilon \equiv v_R h_{\nu}
\eeq
The $R_P$ violating
parameters $\epsilon_i$ and $v_i$ violate lepton number, inducing
a non-zero mass for only one neutrino, which could be considered
to be the the $\nu_{\tau}$. 
The $\nu_e$ and $\nu_{\mu}$
remain massless in first approximation.  As we will explain below, 
they acquire 
masses from supersymmetric loops \cite{numass,ralf} that are typically
smaller than the tree level mass.

\begin{figure}
\begin{picture}(0,8)
\put(0,0){\includegraphics[width=7cm]{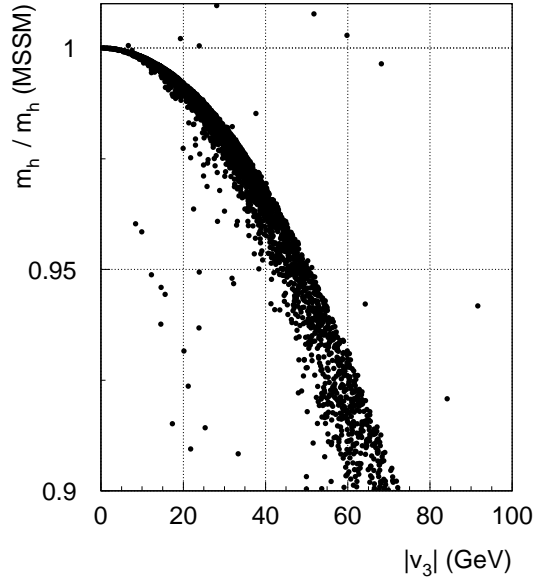}}
\end{picture}
\vspace{-10mm}
\caption{Ratio of the lightest CP-even Higgs
boson mass $m_h$ in the
$\epsilon$--model and in the MSSM  as a function of
$v_3$.}
\label{fig1}
\end{figure}

The model has the MSSM as a limit. This can be illustrated in
Figure~\ref{fig1} where we show the ratio of the lightest CP-even Higgs
boson mass $m_h$ in the
$\epsilon$--model and in the MSSM  as a function of
$v_3$. 
Many other results concerning this model and the implications for
physics at the accelerators can be found in ref.~\cite{e3others,chaHiggsEps}.

\section{Some Results in the Bilinear $\Slash{R}_P$ model}

\subsection{Radiative Breaking: The minimal case}

At $Q = M_{GUT}$ we assume the standard minimal supergravity
unifications assumptions given in Eq.~(\ref{SUGRA}).
In order to determine the values of the Yukawa couplings and of the
soft breaking scalar masses at low energies we first run the RGE's from
the unification scale $M_{GUT} \sim 10^{16}$ GeV down to the weak
scale. We randomly give values at the unification scale
for the parameters of the theory. 
\beq
\begin{array}{ccccc}
10^{-2} & \leq &{h^2_t}_{GUT} / 4\pi & \leq&1 \cr
10^{-5} & \leq &{h^2_b}_{GUT} / 4\pi & \leq&1 \cr
-3&\leq&A/m_0&\leq&3 \cr
0&\leq&\mu^2_{GUT}/m_0^2&\leq&10 \cr
0&\leq&M_{1/2}/m_0&\leq&5 \cr
10^{-2} &\leq& {\epsilon^2_i}_{GUT}/m_0^2 &\leq& 10\cr 
\end{array}
\eeq
The values of $h_{e}^{GUT},h_{\mu}^{GUT},h_{\tau}^{GUT}$ are 
defined in such a way
that we get the charged lepton  masses correctly. 
As the charginos mix with the 
leptons, through a mass matrix given by
\beq
{\cal M}_C=\left[
\matrix{ M_C & A \cr
\vb{18}
B & M_L}
\right]
\eeq
where $M_C$ is the usual MSSM chargino mass matrix,
\beq
M_C=\left[
\matrix{ M & {\textstyle{1\over{\sqrt{2}}}}gv_u\cr
\vb{18}
{\textstyle{1\over{\sqrt{2}}}}gv_d & \mu}
\right]
\eeq
$M_L$ is the lepton mass matrix, that we consider diagonal,
\beq
M_L=\left[
\matrix{ {\textstyle{1\over{\sqrt{2}}}}h_{E_{11}}v_d & 0 & 0 \cr
\vb{18}
0& {\textstyle{1\over{\sqrt{2}}}}h_{E_{22}}v_d & 0 \cr
\vb{18}
0 &0 &{\textstyle{1\over{\sqrt{2}}}}h_{E_{33}}v_d}
\right]
\eeq
and $A$ and $B$ are matrices that are non zero due to the violation of
$R_P$ and are given by
\beq
A^T=\hskip -3pt\left[
\matrix{ -\half h_{E_{11}}v_1 & 0\cr
\vb{18}
-\half h_{E_{22}}v_1 & 0\cr
\vb{18}
-\half h_{E_{33}}v_3 & 0}
\right]
B=\hskip -3pt\left[
\matrix{ \half gv_3 & -\epsilon_1\cr
\vb{18}
\half gv_3 & -\epsilon_2\cr
\vb{18}
\half gv_3 & -\epsilon_3}
\right]
\eeq
We used~\cite{marco} an iterative procedure 
to accomplish that the three lightest
eigenvalues of ${\cal M}_C$ are in agreement with  
the experimental masses of the leptons.
After running the RGE we have a 
complete set of parameters, Yukawa couplings and soft-breaking masses 
$m^2_i(RGE)$ to study the minimization. 
This is done by the following method: we solve the minimization
equations for the soft masses squared. This is easy because those
equations are linear on the soft masses squared. The values obtained
in this way, that we call $m^2_i$ are not equal to the values
$m^2_i(RGE)$ that we got via RGE. To achieve equality we define a function
\beq
\eta= max \left(  \frac{\ds m^2_i}{\ds m^2_i(RGE)},
\frac{\ds m^2_i(RGE)}{\ds m^2_i}
\right) \quad \forall i 
\eeq
with the obvious property that
\beq
\eta \ge 1
\eeq
Then we adjust the parameters to minimize $\eta$.
Before we end this section let us discuss the counting of free
parameters in this model and in the minimal N=1 supergravity unified
version of the MSSM. In \Table{table1} we show this counting for the
MSSM and in \Table{table2} for the $\epsilon$--model.
Finally, we note that in either case, the sign of the mixing parameter
$\mu$ is physical and has to be taken into account.
\begin{table}
\caption{Counting of free parameters in MSSM}
\begin{tabular}{ccc}\hline
Parameters  
\hskip -9pt&\hskip -9pt 
Conditions 
\hskip -9pt&\hskip -9pt 
Free Parameters \cr \hline
$h_t$, $h_b$, $h_{\tau}$
\hskip -9pt&\hskip -9pt 
$m_W$, $m_t$
\hskip -9pt&\hskip -9pt 
 $\tan\beta$ \cr 
$v_d$, $v_u$,$M_{1/2}$ 
\hskip -9pt&\hskip -9pt 
$m_b$, $m_{\tau}$ 
\hskip -9pt&\hskip -9pt 
 2 Extra \cr 
 $m_0$, $A$, $\mu$
\hskip -9pt&\hskip -9pt 
$t_i=0$, $i=1,2$
\hskip -9pt&\hskip -9pt 
({\it e.g.} $m_h$, $m_A$)\cr \hline
Total = 9
\hskip -9pt&\hskip -9pt 
Total = 6 
\hskip -9pt&\hskip -9pt 
Total = 3\cr\hline
\end{tabular}
\label{table1}
\end{table}

\begin{table}
\caption{Counting of free parameters in our model}
\begin{tabular}{ccc}\hline
Parameters 
\hskip -9pt&\hskip -9pt 
 Conditions 
\hskip -9pt&\hskip -9pt 
 Free Parameters \cr \hline
$h_t$, $h_b$, $h_{\tau}$
\hskip -9pt&\hskip -9pt 
$m_W$, $m_t$ 
\hskip -9pt&\hskip -9pt 
 $\tan\beta$, $\epsilon_i$ \cr 
$v_d$, $v_u$, $M_{1/2}$
\hskip -9pt&\hskip -9pt 
$m_b$, $m_{\tau}$ 
\hskip -9pt&\hskip -9pt  \cr 
$m_0$,$A$, $\mu$
\hskip -9pt&\hskip -9pt 
$t_i=0$
\hskip -9pt&\hskip -9pt 
 2 Extra \cr 
$v_i$, $\epsilon_i$
\hskip -9pt&\hskip -9pt 
($i=1,\ldots,5$)
\hskip -9pt&\hskip -9pt 
 ({\it e.g.} $m_h$, $m_A$)\cr \hline
Total = 15
\hskip -9pt&\hskip -9pt 
Total = 9 
\hskip -9pt&\hskip -9pt 
Total = 6\cr\hline
\end{tabular}
\label{table2}
\end{table}

\subsection{Gauge and Yukawa Unification in the $\epsilon$ model}

\begin{figure}[ht]
\begin{center}
\includegraphics[width=7cm]{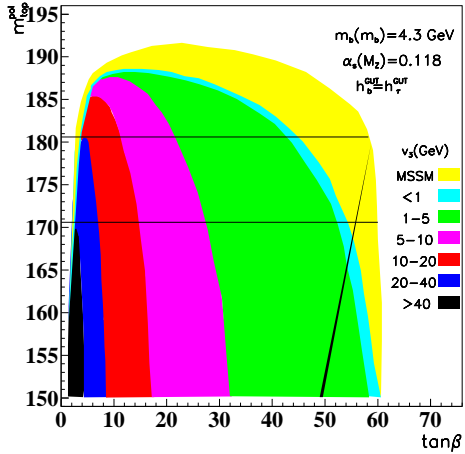}
\end{center}
\vspace{-10mm}
\caption{Top quark mass as a function of $\tan\beta$ for different
values of the R--Parity violating parameter $v_3$.}
\label{aretop}
\end{figure}

There is a strong motivation to consider GUT theories where {\it
both} gauge and Yukawa unification can achieved. This is because 
besides achieving gauge coupling unification,
GUT theories also reduce the number of free parameters in the Yukawa
sector and this is normally a desirable feature. The situation with
respect to the MSSM can be summarized as follows.
In $SU(5)$ models, $h_b=h_{\tau}$ at $M_{GUT}$.
The predicted ratio $m_b/m_{\tau}$ at $M_{WEAK}$ agrees with
experiments. 
A relation between $m_{top}$ and
$\tan\beta$ is predicted. Two solutions are possible: low and high 
$\tan\beta$.
In $SO(10)$ and $E_6$ models $h_t=h_b=h_{\tau}$ at $M_{GUT}$.
In this case, only the large
$\tan\beta$ solution survives.
We have shown~\cite{YukUnifBRpV} that the $\epsilon$--model allows $b-\tau$ Yukawa unification for
any value of $\tan\beta$ and satisfying perturbativity of the
couplings.  We also find the $t-b-\tau$ Yukawa unification 
easier to achieve than in the MSSM, occurring in a 
wider high $\tan\beta$ region. This is shown in
Fig.~{\ref{aretop} where we plot the top quark mass as a
function of $\tan\beta$ for different
values of the R--Parity violating parameter $v_3$. Bottom quark and
tau lepton Yukawa couplings are unified at $M_{GUT}$. The horizontal
lines correspond to the $1\sigma$ experimental $m_t$
determination. Points with $t-b-\tau$ unification lie in the diagonal
band at high $\tan\beta$ values. We have taken $M_{SUSY}=m_t$.}

\subsection{On $\alpha_s(M_Z)$ versus $\sin^2 \theta_W (M_Z)$}

\begin{figure}[ht]
\begin{center}
\includegraphics[width=7cm]{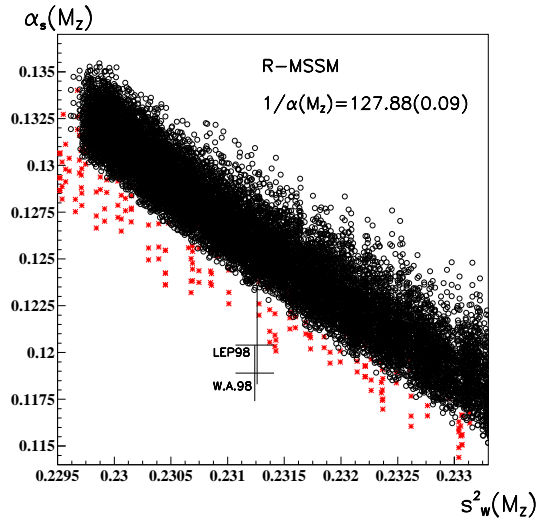}
\end{center}
\vspace{-10mm}
\caption{$\alpha_s(M_Z)$ versus $\hat s_Z$ for the MSSM (circles) and
for the bilinear $\not{\!\!\!R}_p$ model (crosses).}
\label{nmssm-sfeliu}
\end{figure}

Recent studies  of gauge coupling unification in the context 
of MSSM agree that using
the experimental values for the electromagnetic coupling and the weak
mixing angle, we get the prediction 
\beq
\alpha_s(M_Z) \sim 0.129 \pm 0.010
\eeq
that it is about 1$\sigma$ larger than indicated by the most recent
world average value 
\beq
\alpha_s(M_Z)^{W.A}=0.1189 \pm 0.0015
\eeq
We have re-considered~\cite{alfas}
 the $\alpha_s$ prediction in the context of the
model with bilinear breaking of R--Parity. We have shown
that in this simplest R--Parity breaking model,
with the same particle content as the MSSM, there appears an
additional negative contribution to $\alpha_s$, which can bring the
theoretical prediction closer to the experimental world average. 
This additional contribution comes from two--loop b--quark Yukawa effects
on the renormalization group equations for $\alpha_s$. Moreover we
have shown that this contribution is typically correlated to the
tau--neutrino mass which is induced by R--Parity breaking and which
controls the R-Parity violating effects. We found that it is possible
to get a 5\% effect on $\alpha_s(M_Z)$ even for light $\nu_{\tau}$
masses. This is shown in Fig.~\ref{nmssm-sfeliu} where we compare the
predictions of $\alpha_s(M_Z)$ in the MSSM and in the bilinear
$\not{\!\!\!R}_p$ model. 

\subsection{Neutrino Physics}

In this model at tree-level only one neutrino picks up a mass via the
mixing with the neutralinos. This result is exact but it can best be
seen in the limit where the \rp parameters are small compared with the
SUSY parameters~\cite{MartinValle},
\beq
\epsilon_i, g v_i, g' v_i \ll M_1,M_2,\mu
\eeq
Then we can write an effective neutrino $3\times3$ matrix (see--saw)
\beq
m_{eff} \!=\! 
\frac{M_1 g^2 \!\!+\!\! M_2 {g'}^2}{4\, \det({\cal M}_{\chi^0})} 
\left(\!\!\begin{array}{ccc}
\Lambda_e^2\!\! &\!\! \Lambda_e \Lambda_\mu
\!\!&\!\! \Lambda_e \Lambda_\tau \\
\Lambda_e \Lambda_\mu \!\!&\!\! \Lambda_\mu^2
\!\!&\!\! \Lambda_\mu \Lambda_\tau \\
\Lambda_e \Lambda_\tau \!\! &\!\! \Lambda_\mu \Lambda_\tau \!\!&\!\! \Lambda_\tau^2
\end{array}\!\!\right)
\eeq
where $\det({\cal M}_{\chi^0})$ is the determinant of the MSSM
neutralino mass matrix and
\beq
\Lambda_i = \mu v_i + v_d \epsilon_i
\eeq
The projective nature of $m_{eff}$ ensures that we get two zero
eigenvalues. The only non--zero is
\beq
m_{\nu} = Tr(m_{eff}) = 
\frac{M_1 g^2 + M_2 {g'}^2}{4\, det({\cal M}_{\chi^0})} 
|{\vec \Lambda}|^2.
\eeq
At 1--loop level the two massless neutrinos get masses. The masses and
mixings can be shown~\cite{numass} 
to be compatible with those needed to solve the
solar and atmospheric neutrino problems. We refer to the talk of
M. Hirsch at this Conference~\cite{martin} for the details. 

\subsection{Results at the Accelerators}

\begin{figure}[ht]
\begin{center}
\includegraphics[angle=90,width=75mm]{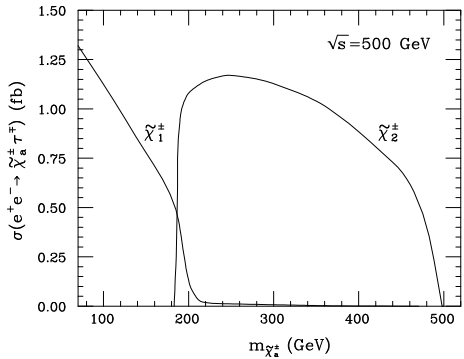}
\end{center} 
\vspace{-15mm}
\caption{Maximum single chargino production cross section as a function 
of the chargino mass at NLC in BRpV. Light and heavy charginos are 
displayed.}
\label{csnlc}
\end{figure}

\begin{figure}[ht]
\begin{center}
\includegraphics[width=70mm]{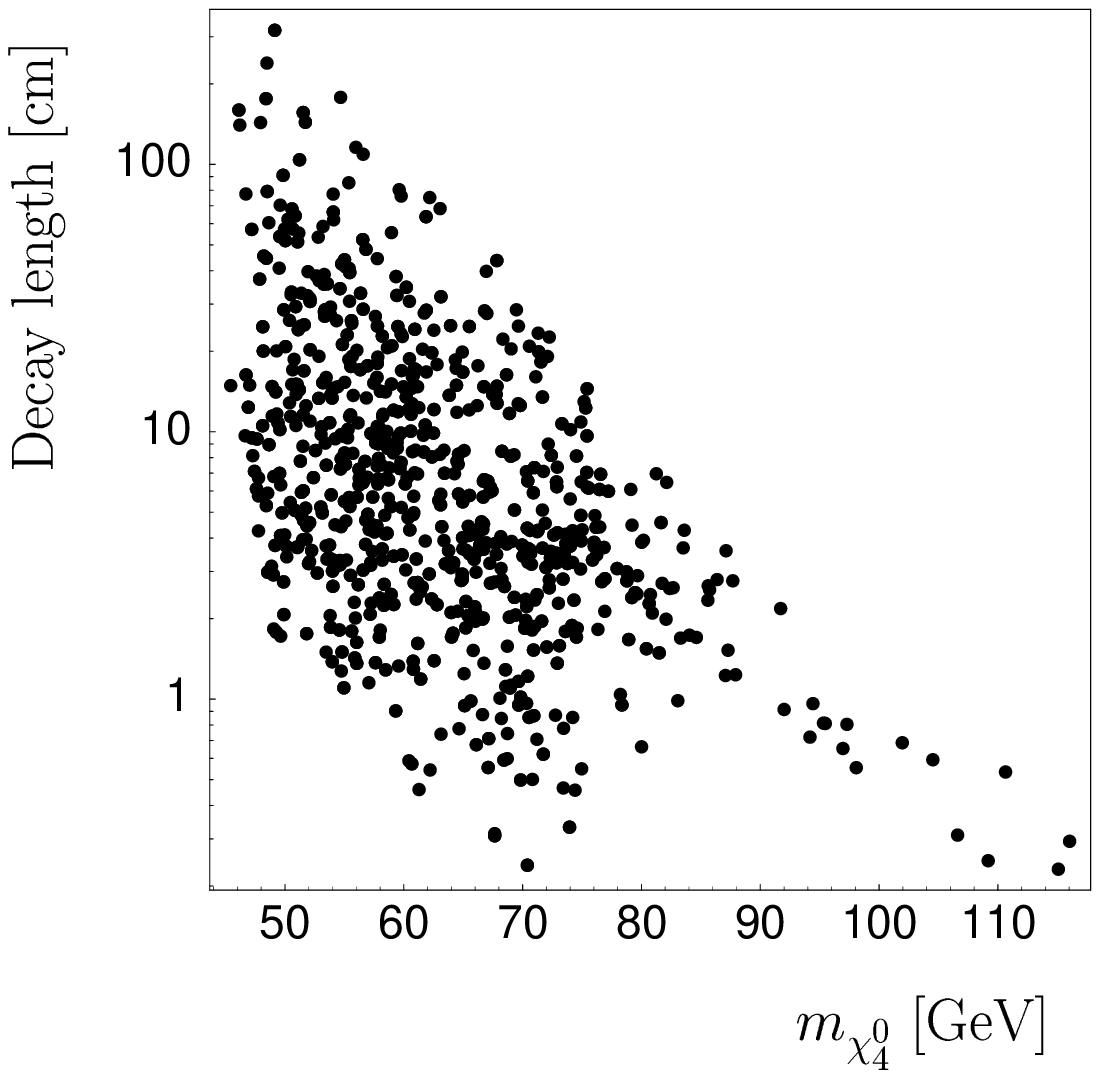}
\end{center}
\vspace{-10mm}
\caption{}
\label{neudecay}
\end{figure}

We have extensivley studied the implications of the BRpV model at the
accelerators~\cite{chaHiggsEps,SBRPAcc}. For instance an important 
prediction of the BRpV model is that the chargino can be
single produced. The prediction for the NLC is shown in Fig.~\ref{csnlc}
Here we will not describe these results any further but we 
emphasize that if  R-parity is violated, the neutralino is 
unstable. As it shown in Fig.~\ref{neudecay} it decays inside the
detector. This is very important because its decays can serve as
probes for the solar neutrino parameters~\cite{martin,porod}.

\section{Conclusions}

We have shown that there is a viable model for {\it SBRP} that 
leads to a very rich phenomenology, both at laboratory experiments, 
and at pre\-sent (LEP) and future (LHC, LNC) accelerators.
In these models the radiative breaking of {\it both} the Gauge
Symmetry and R-Parity can be achieved.
Most of these phenomenology can be described by an {\it effective} model
with {\it explicit} R--Parity violation.
This model has many definite predictions that are different from the
MSSM. In particular $b-\tau$ can be achieved for any value of $\tan
\beta$ and we get a better prediction for $\alpha_s(M_Z)$ then in the
MSSM. 
We have calculated the {\it one--loop} corrected {\it masses}
and {\it mixings} for the neutrinos and we get that these have the
correct features to explain both the solar and atmospheric neutrino
anomalies. 
We emphasize that the 
{\it lightest neutralino decays inside the detectors},
thus leading to a very different phenomenology than the MSSM. 
In particular the neutrino parameters can be tested at the
accelerators.

\end{document}